\documentstyle[preprint,aps,eqsecnum,epsfig]{revtex}

\def\gtrsim{\mathrel{\mathpalette\vereq>}}

\def\doeack{\footnote{Work supported by the Department of Energy,
                     contract DE--AC03--76SF00515.}}
\def\SLAC{Stanford Linear Accelerator Center\\
    Stanford University\\ Stanford, CA 94309}

\def\Journal#1#2#3#4{{#1} {\bf#2}, {#3} {(#4)}}

\def\PLB{{ Phys. Lett.}  B}
\def\PRL{ Phys. Rev. Lett.}
\def\PRD{{ Phys. Rev.} D}

\def\EPJC{{Eur. Phys. J.} C}

\begin{document}
\draft
\preprint{\vbox{\hbox{SLAC-PUB-8309}   
                \hbox{December 1999}
                \hbox{}}}
             
\title{\vbox{\vskip 0.truecm} Yukawa Hierarchies from Split Fermions in Extra Dimensions}

\author{Eugene A. Mirabelli and Martin Schmaltz\doeack }
\address{\vbox{\vskip 0.truecm} \SLAC \\ \vbox{\vskip 0.truecm} 
         {\tt gino@SLAC.stanford.edu, schmaltz@SLAC.stanford.edu}}
\maketitle
\vskip.4in
\begin{abstract}
We explore a mechanism for generating the Yukawa hierarchies by
displacing the left- and right-handed components of the Standard
Model fermions in a higher-dimensional space. We find a unique
configuration of displacements which fits all quark and lepton
masses and mixing angles, with a prediction for the strange quark
mass $m_s^{\overline{MS}}(2 \  GeV) \approx (1.19) \times (V_{ub}
V_{cb}/V_{us})^{1/2} \times m_b^{\overline{MS}}(m_b) \approx 120 \  MeV$.
\end{abstract}

\pacs{}

\section{Introduction}
The parameters of the standard model contain a mystery: why do the Yukawa couplings vary over so many orders of magnitude?  The up quark Yukawa coupling is of order unity, which we would na\"{\i}vely expect, but the other couplings are distributed hierarchically over several orders of magnitude below this.  
Why are most of the standard model fermions so light? 

Traditional explanations of this Yukawa hierarchy invoke new symmetries, but
in \cite{arkaniHamedSchmaltz} Arkani-Hamed and Schmaltz suggested an
explanation which does not require imposing new symmetries.  They suggest that
the hierarchies present in the Yukawa couplings of the standard model can be
explained as a result of the slight displacement of the standard model field
wavefunctions inside a four-dimensional domain wall in a higher-dimensional
space.  The effective four-dimensional Yukawa coupling is a product of the
fundamental higher-dimensional Yukawa coupling and the overlap of the field
wavefunctions.  If the wavefunctions are highly peaked, a small relative shift
between wavefunctions leads to a large suppression of the effective Yukawa
coupling through the smallness of the overlap of the wavefunctions.  Even
with a single $O(1)$ higher-dimensional Yukawa coupling, one could produce
Yukawa hierarchies through the relative displacements of the wavefunction
peaks.

Ref. \cite{arkaniHamedSchmaltz} also provides a natural, field-theoretic
mechanism for producing such displacements, and demonstrates how it can lead
to the appearance of a small Yukawa coupling.  However, the discussion in
\cite{arkaniHamedSchmaltz} is very general, which raises the question of
whether a realistic model can actually be constructed, or whether there are
hidden constraints and relations among the masses and mixing angles of the
standard model that preclude their generation from a set of displaced
wavefunctions.

In this paper we seek to address this question. We find that while it is
easy to arrange two- (or higher) dimensional arrays of wavefunctions which
reproduce the parameters of the standard model, it is difficult (but not
impossible!) to do so in a minimal one-dimensional version; we find essentially
one possible configuration of wavefunctions. This configuration has a definite prediction, which may be confirmed or
falsified by more accurate determinations of the strange quark mass and $V_{ub}$.  For alternative extra-dimensional approaches to flavor see \cite{flavor}.

In section \ref{producingHierarchiesWithExponentiallySmallOverlaps}, we
summarize the mechanism of \cite{arkaniHamedSchmaltz}, and in section
\ref{theModel} we summarize the outline of a model suggested in that paper. 
In section \ref{leptonSectorOfTheModel} we demonstrate the mechanism for the
simple case of the lepton sector, and in section \ref{quarkSectorOfTheModel}
we outline the challenges present when trying to model the quark sector.  In
section \ref{searchingParameterSpace} we describe our efforts to find a set of
locations for the quark wavefunctions, and in section \ref{analysisOfSolution}
we analyze the (essentially) unique set, which leads to the prediction for the
strange quark mass $m_s \approx (1.19) \times m_b (V_{ub} V_{cb}/V_{us})^{1/2}
\approx 120 \ MeV $.  We briefly conclude in section \ref{gaussConclusion}. 
Appendix \ref{derivationOfZeroMode} contains a derivation of the form of the
zero mode for a minimal five-dimensional spinor coupled to a scalar with a
domain wall profile expectation value, and appendix
\ref{standardModelParameters} contains the experimentally allowed masses and
mixings we used, as well as the renormalization multipliers used to evaluate
those parameters at a common scale.  

\section{Producing Hierarchies With Exponentially Small Overlaps}
\label{producingHierarchiesWithExponentiallySmallOverlaps}
In this section, we summarize the mechanism of \cite{arkaniHamedSchmaltz} for
producing hierarchies between mass matrix elements.   We start by considering a
theory in which the standard model fermions are trapped in a four-dimensional
membrane in a five-dimensional space.  Although non-perturbative effects in
string/M theory can be used to achieve such localizations, one can also study
models in which the localization occurs in the context of an effective field
theory.  

\begin{figure}
\begin{center}
\leavevmode{\epsfysize=3.0truein \epsfbox{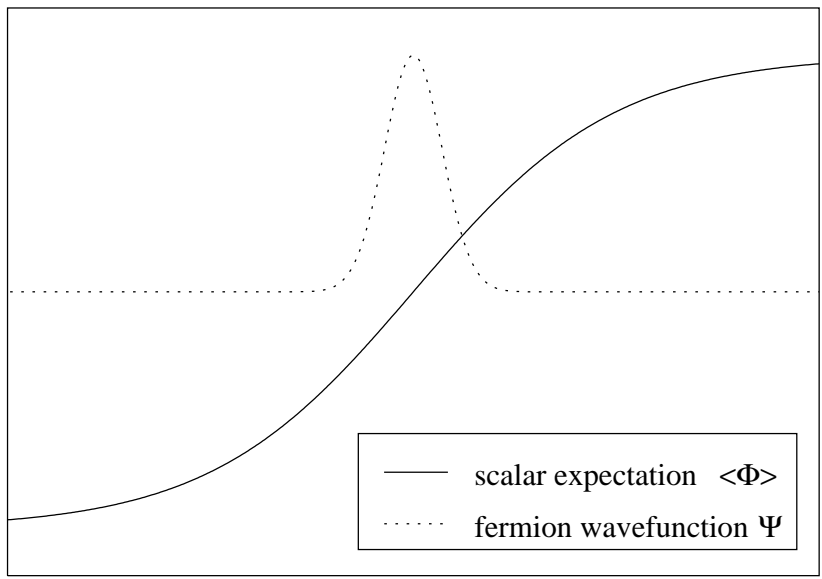}}
\end{center}
\caption{A chiral fermion trapped on a domain wall.}
\label{tanhGaussFigure}
\end{figure}

One concrete field-theory method of achieving such a localization is to couple
a massless five-dimensional fermion $\Psi$ to a five-dimensional scalar field
$\Phi$ through a Yukawa coupling term $\sim \int {\rm d}^5x
\Phi\overline{\Psi}\Psi$ in the Lagrangian, and then to include an effective
potential for $\Phi$ with two or more isolated vacua, such as a potential
$\sim (\Phi^2- v^2)^2$, so that it attains a position-dependent vacuum
expectation value with a domain-wall profile in the extra dimension,
interpolating between regions with different choices of vacuum (see figure
\ref{tanhGaussFigure}).  

The fermion then acquires a zero-mode localized near the zero-crossing
of $\Phi$.   Heuristically, one may think of the Yukawa coupling to the
position-dependent scalar expectation value as giving the fermion a
position-dependent mass which greatly suppresses field fluctuations far from
the domain wall at $x_5 = 0$ while leaving fluctuations at the origin massless.
Although a single minimal spinor in five-dimensional space decomposes into a
left-right mirror pair of chiral fermions in four-dimensional language, only
one chirality zero-mode is trapped on the wall.  

In section \ref{derivationOfZeroMode} we derive the zero mode solution in the approximation in which the scalar field profile is a linear function of the extra dimension $\Phi = 2 \mu^2 x_5$ (which is valid for points close to the center of the domain wall) and find that zero mode solutions for the fermions have Gaussian profiles in the extra dimension centered on the zero of $\Phi(x_5)$:
\begin{equation}
\Psi(x)=A \ e^{-\mu^2{x_5}^2}  \times \psi(x_0, x_1, x_2, x_3) ,
\label{zeroModeSolution}
\end{equation}
where $\psi$ is a canonically normalized massless left-handed four-dimensional
fermion field, $A = \mu^{1/2}(\pi/2)^{-1/4}$ is a normalization constant, and
$\mu$ is related to the slope of the scalar field profile by $ \mu^2 =
\partial_5\langle \Phi \rangle / 2$. Higher modes, which come in mirror pairs,
have minimum energy $\approx \mu$.

\begin{figure}
\begin{center}
\leavevmode{\epsfysize=3.00truein \epsfbox{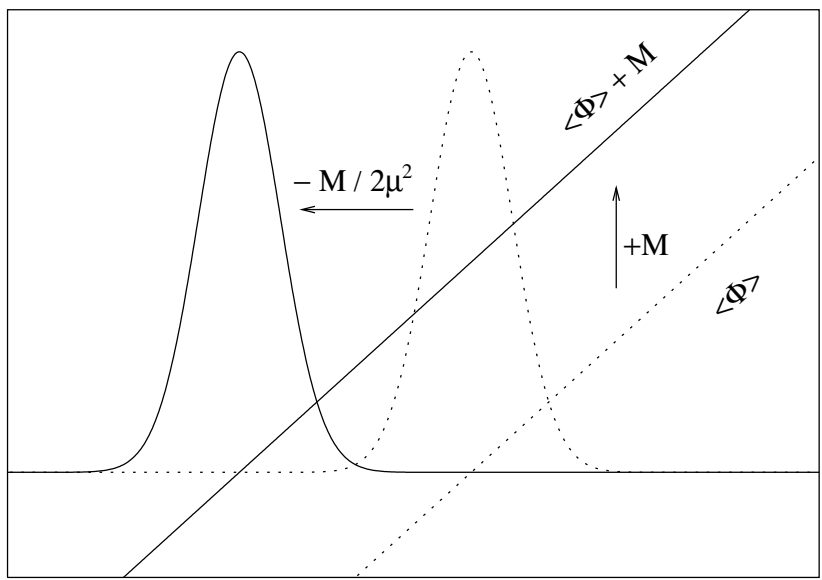}}
\end{center}
\caption{A mass term shifts the location of the domain wall.}
\label{shiftFigure}
\end{figure}

In \cite{arkaniHamedSchmaltz}, it was pointed out that adding a
five-dimensional fermion mass term $ \int {\rm d}^5x M\overline{\Psi}\Psi$ to
the above system acts like an effective shift of the scalar field $\Phi
\rightarrow \Phi + M$ as far as the fermion is concerned, which (in the
approximation in which the scalar field profile is linear) leads to an
effective shift of the location where the scalar has a zero expectation value,
i.e., a shift in the effective location of the domain wall as seen by the
fermion (see figure \ref{shiftFigure}) so that it ends up localized around
$\ell_5 = - M / 2 \mu^2$.    If different fermions have different masses $M_i$,
they each end up localized around different locations $\ell_{(i)5} = - M_i / 2
\mu^2$, each fermion being centered on a different ``effective'' zero of the
same scalar field.

We assume a generic five-dimensional Yukawa coupling between two fermions and a Higgs scalar field $H$, which we take for simplicity to have a constant profile in the extra dimension.  This leads, upon substitution of the zero mode solution from equation (\ref{zeroModeSolution}) and integration over the extra dimension, to a four dimensional coupling modulated by the mutual overlap of the fermion wavefunctions:
\begin{eqnarray}
\ {\cal L}_{Yukawa} & =  & \int {\rm d}^5x\ \sqrt{L}\ \kappa \ \ H \overline{\Psi}_1 \Psi_2  \\
\ & = & \int {\rm d}^5x \ \sqrt{L}\ \kappa  \ \ \left({1 \over \sqrt{L}} \ h \right)  \left(  A\ e^{-\mu^2{(x_5-\ell_1)}^2}  \psi_1 \right) \left(  A\ e^{-\mu^2{(x_5-\ell_2)}^2} \psi_2 \right)  \\
\ & = & \int {\rm d}^4x \ \left( \int {\rm d}x_5\ \kappa  A\ e^{-\mu^2{(x_5-\ell_1)}^2}  A\ e^{-\mu^2{(x_5-\ell_2)}^2}  \right)   h  \psi_1 \psi_2\\
\ & = & \int {\rm d}^4x \ (e^{-\frac{1}{2}\mu^2{(  \ell_1 - \ell_2 )}^2}\kappa)  \ h  \psi_1\psi_2 \\
\ & = &\int {\rm d}^4x \ \lambda \  h  \psi_1 \psi_2,
\end{eqnarray}
where we have inserted a factor of the square root of the domain wall width $\sqrt{L}$ to render the coupling constant $\kappa$ dimensionless.  The effective four-dimensional Yukawa coupling:
\begin{equation}
\lambda = e^{-\frac{1}{2} \mu^2{(\ell_1 - \ell_2)}^2}\kappa,
\label{couplingFromDisplacement}
\end{equation}
depends on the mutual overlap of the zero modes of the fermions through the relative displacement of their peaks $| \ell_1 - \ell_2|$.  
If the fermions are more than a few Gaussian widths $\mu^{-1}$ from each other then an order-one coefficient for the five-dimensional Yukawa coupling $\kappa$ results in an exponentially small four-dimensional Yukawa coupling $\lambda$.  
Setting the Higgs field to its expectation value and generalizing to many generations gives a mass matrix element:
\begin{eqnarray}
m_{ij}& = &  \left( \langle h \rangle \ \kappa \right) \  e^{-\frac{1}{2} \mu^2{(\ell_i - \ell_j)}^2} \nonumber \\
& = & \rho \ e^{-\frac{1}{2} \mu^2{(\ell_i - \ell_j)}^2},
\label{massFromDisplacement}
\end{eqnarray}
where $ \rho \equiv \langle h \rangle \ \kappa $.  
In this paper we seek to obtain the hierarchical pattern of masses and mixings in the standard model as a result of a particular configuration of wavefunction locations with relative separations of a few Gaussian widths $\mu^{-1}$.

It is useful to invert equation (\ref{massFromDisplacement}) to find the relative displacements associated with a mass matrix:
\begin{equation}
|\ell_i - \ell_j| \  = \ \mu^{-1} \sqrt{-2 \ {\rm log}( m_{ij} / \rho)} \ \equiv \ p(m_{ij}).
\label{displacementFromMass}
\end{equation}
To produce a particular mass matrix $m_{ij}$, we need to find a set of wavefunction locations which have relative displacements $|\ell_i - \ell_j| $ related to the mass matrix by the above formula.  The function $\mu^{-1} \sqrt{-2 \ {\rm log}( m_{ij} / \rho)}$ appears often in expressions to follow; we refer to this function as $p(m_{ij})$.

\section{The Model}
\label{theModel}

We now summarize the model outlined in \cite{arkaniHamedSchmaltz}. By the
standard model gauge symmetry, all fields in a given gauge multiplet have the same effective five dimensional fermion mass $M_i$ and thus have wavefunctions peaked at the same location in the extra
dimension.   This means that there are fifteen wavefunction locations $q_i,
u_i, d_i, l_i$ and $e_i$ for the left-handed standard model fields $Q_i, U^c_i,
D^c_i, L_i$ and $E^c_i$ (with $i=1,2,3$). The fermions are all taken to have
equal couplings to the domain wall scalar $\Phi$ (meaning their Gaussian
wavefunctions all have equal widths), and so as not to add extra complication,
we take the Higgs field $H$ to have an expectation value which is constant
throughout the domain wall, and we take the effective five dimensional Yukawa
couplings between the fermions and the Higgs to have a single
magnitude.   Although these restrictions could be relaxed, the structure
already present proves sufficient to generate the standard model masses and
mixings.   We defer a discussion of the CP violating complex phase of the CKM
matrix to later work, and for the present time study only real Yukawa
matrices.  

The analysis of Arkani-Hamed and Schmaltz constrains the scales present in the
model.   For the description to make sense, we need the wall thickness $L$ to
be larger than the Gaussian width $\mu^{-1}$, which in turn should be larger
than the length scale of the ultraviolet cutoff $1/M_*$.   For the four
dimensional effective top Yukawa coupling $\lambda_t$ to be perturbative at
$M_*$ we must further constrain  $N \lambda_t^2/16\pi^2  < 1 $ where $N =
(M_*L)/2\pi$ is the number of Kaluza-Klein modes below the cutoff.   If a field
theory description is to apply throughout the wall, we must have $\Phi(L/2)
\sim  \mu^2 L  < M_*$.  To summarize: 
\begin{eqnarray}
L^{-1} & < & \mu  < M_* \label{firstMagnitudeConstraint} \\
M_* / L^{-1} & < & 32 \pi^3 / \lambda_t^2  \approx 1000 \\
\mu / L^{-1}  & < & \sqrt{ M_* / L^{-1}} \approx 30 \label{lastMagnitudeConstraint}.
\end{eqnarray}

Note that these inequalities only constrain the ratios of the mass scales.  The overall mass scale corresponds to a free parameter of the model.  It would be most interesting if this overall mass scale were low, so that the extra-dimensional physics would be accessible to experiment.  Constraints on the size of the extra dimension $L$ come from direct searches for Kaluza-Klein excitations \cite{KK} and precision electroweak data \cite{electroweakData}.  Constraints on split fermions were discussed in \cite{arkaniHamedSchmaltz,yuval}.  However, the most important constraints come from considering flavor changing neutral currents mediated by Kaluza-Klein gauge bosons \cite{FCNC}, and lead to the bound 
\begin{equation}
L^{-1}   \gtrsim   100 \  {\rm TeV} \ . 
\end{equation}

In equation (\ref{massFromDisplacement}) the mass matrix element is bounded from
above by $\rho$, with the maximal value achieved when the fermion wavefunctions
are localized around the same point.   This means that to generate the top mass
from a single matrix element the condition $\rho > m_t$ is required. On the
other hand, naturalness and perturbativity of couplings lead us to want $\rho$
as small as possible.   For concreteness, we choose $\rho$ to be $1.5 \  m_t$. 
Our model works for any $\rho \geq m_t$, with larger $\rho$ values requiring
greater suppressions from the overlaps, and thus requiring larger separations
between wavefunctions.  The mass prediction of equation (\ref{lowMassRelation})
is independent of $\rho$, since the wavefunction positions are always chosen to
reproduce the same mass matrices.

Before trying to reproduce the masses of the standard model, we first run the masses of the standard model to a common scale $m_t$ using the scaling factors found in the appendix (the running of the CKM elements to that scale is small compared to the experimental uncertainty in their magnitudes).  

\section{Lepton sector of the model}
\label{leptonSectorOfTheModel}

We begin our efforts to reproduce the standard model masses and mixings by finding a set of wavefunction positions for the $L_i$ and $E_i^c$ fields which reproduce the observed charged lepton masses.  This is easy, since we are only trying to fit three parameters and we may adjust five parameters (six locations, minus one freedom to shift the locations by an overall displacement since only relative locations matter in equation (\ref{massFromDisplacement}) ).  
Assuming a simple diagonal texture for the lepton mass matrix:
\begin{equation}
{\bf m_e} = \pmatrix{m_e&0&0\cr0&m_{\mu}&0\cr0&0&m_{\tau}}, \label{electronTexture}
\end{equation} 
leads, using equation (\ref{displacementFromMass}) to a matrix of relative distances:
\begin{equation}
|l_i - e_j|  = \mu^{-1} \pmatrix{5.1275 &(far)&(far)\cr(far)&3.9475&(far)\cr(far)&(far)&3.1498 }.
\end{equation}

We have three constraints on the locations, with the only other constraints
being that the relative displacements indicated by $(far)$  be great enough to
lead to negligible off-diagonal matrix elements. In fact, it turns out that
little care is necessary to insure that the off-diagonal elements are
negligible; one need only insure that, for a given $L$ field, the closest
$E^{c}$ field is roughly one width closer than the second-closest $E^{c}$
field. It proves to be a general property of this model, due to the exponential
in equation (\ref{massFromDisplacement}), that if one wavefunction is near two
others and the two others are not the same distance away then the overlap with
the farther one will often be negligible compared to the overlap with the
closer one.   In other words, it is easy to generate elements of the Yukawa
matrix that are effectively zero.

\begin{figure}
\begin{center}
\leavevmode{\epsfxsize=6.00truein \epsfbox{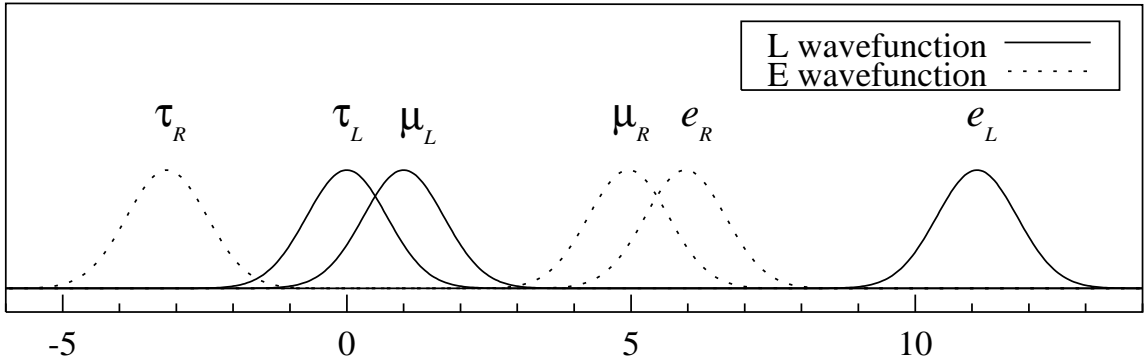}}

\end{center}
\caption{The locations of the lepton wavefunctions.}
\label{oneDimensionalLeptonFigure}
\end{figure}

One possible set of locations for the lepton fields is:
\begin{equation}
 l_i   =   \mu^{-1} \pmatrix{11.075 \cr 1.0 \cr 0.0} , \ \ \ \ \ \ \ \ \ \ \ \ \ \ \ e_i  =  \mu^{-1} \pmatrix{5.9475  \cr 4.9475  \cr -3.1498 } . \label{leptonSolution} 
\end{equation}
This configuration is depicted graphically in figure
\ref{oneDimensionalLeptonFigure}. This is by no means the only set of
wavefunction locations that reproduce the $e$, $\mu$, and $\tau$ masses.   For
instance, the mass spectrum is essentially unchanged if one  moves the $L_i$
further apart while keeping each $E_i^c$ wavefunction the same distance from
its partner $L_i$.  Also, one can
consider textures differing from that of equation
(\ref{electronTexture})---since lepton mixing angles are unobservable, all that
is necessary is that the eigenvalues of the mass matrix equal the lepton
masses.  Different textures would lead to different sets of wavefunction
locations.

\section{Quark sector of the model}
\label{quarkSectorOfTheModel}

We now outline the challenges present when trying to model the quark sector.  
It is more difficult than the lepton sector because one has to fit nine
observables (six masses and three mixing angles) by adjusting eight parameters
(the nine locations of the $Q_i$, $U_i^c$, and $D_i^c$, minus one freedom to
shift the locations by an overall displacement).

In fact, equation (\ref{massFromDisplacement}) determines the mass matrices
$\bf{m}_u$ and $\bf{m}_d$.   To obtain the physical masses and CKM elements
from the mass matrices we must perform unitary redefinitions of the
$Q_{(I=+1/2)i}$, $Q_{(I=-1/2)i}$, $U_i^c$ and $D_i^c$ fields to a basis where
the mass matrices are diagonal.   The masses can be read off the diagonal
entries, and the CKM matrix appears in the weak interaction terms as a result
of the mismatch between the redefinitions of the isospin-up and isospin-down
components of the quark doublets.  

A useful method of computing the parameters is to calculate:
\begin{eqnarray}
\ & \bf{m}_u \bf{m}^\dagger_u = \bf{U}_u \pmatrix{m^2_u&0&0\cr0&m^2_c&0\cr0&0&m^2_t\cr} \bf{U}^\dagger_u \label{firstComputePhysicalParameters} \\
\ & \bf{m}_d \bf{m}^\dagger_d =\bf{U}_d \pmatrix{m^2_d&0&0\cr0&m^2_s&0\cr0&0&m^2_b\cr} \bf{U}^\dagger_d\\
\nonumber \\
\ & \bf{V}_{CKM} = \bf{U}^\dagger_u \bf{U}_d. \label{lastComputePhysicalParameters}
\end{eqnarray}
The eigenvalues of the hermitian squares of the up(down) mass  matrices give the squares of the masses of the up(down) type quarks, and the products of the eigenvectors of the hermitian square of the up mass matrix with the eigenvectors of the hermitian square of the down mass matrix give the CKM matrix elements.  
Ignoring the CP violating complex phase, it is sufficient to match the magnitudes of the three components of the CKM matrix above the diagonal to match the entire matrix.

However, we seek to do more than simply reproduce the standard model masses and
mixing angles---we seek to reproduce the hierarchies present in a natural way,
as a result of the exponential in equation (\ref{massFromDisplacement}).   This
excludes certain mass matrix textures from consideration.

For example, consider the mass matrices:
\begin{eqnarray}
\pmatrix{0&0\cr 0&1} \ & \  {\rm and } \ &  \ \pmatrix{.5&.5\cr .5&.5}. \
\end{eqnarray}
Both yield a spectrum with a hierarchy (in fact, they each yield a spectrum
containing one massless and one massive particle) but in the first matrix the
hierarchy is a result of the hierarchy between the last element and all the
other elements, which equation (\ref{massFromDisplacement}) easily produces,
while in the second matrix the hierarchy is a result of a delicate cancellation
between all the matrix elements, which would be exponentially fine-tuned if
these elements came from equation (\ref{massFromDisplacement}).   Although these
fine tunings could possibly be explained by adding further symmetries, our
philosophy is to seek a set of generic-looking positions which lead to small
masses and mixings as a result of small numbers coming from the exponential in
(\ref{massFromDisplacement}).   We only allow small numbers to come from the
product of small numbers, but not from a cancellation or sum.   One might be
tempted to allow cancellations or sums from numbers which are of the same
order-of-magnitude as the quantity they cancel or sum to produce, since it does
not seem like fine-tuning to have $O(\epsilon^n) \pm  O(\epsilon^n) =
O(\epsilon^n)$, but we avoid doing this here because in this model it would
usually involve an unexplained coincidence---generically any two elements of
the mass matrix resulting from equation (\ref{massFromDisplacement}) have vastly
different orders of magnitude and it takes a strange coincidence or a fine
tuning for two elements to have an $O(1)$ ratio.   An exception to this is the
third generation: the masses of the top and bottom approach the upper-bound
scale $\rho$, and as we approach this scale such coincidences become less
fine-tuned\footnote{The fine tuning  $\partial m_{ij} / \partial
\ell_j \rightarrow 0$ as $|\ell_i - \ell_j| \rightarrow 0$, which happens when 
$m_{ij} \rightarrow \rho$.  For $|\ell_i - \ell_j|$ larger than a few widths
$\mu^{-1}$, that is, for $m_{ij}$ below the mass scale of the third generation,
$\partial m_{ij} / \partial \ell_j$ grows rapidly.}, so we allow models which
produce a mass parameter $\sim m_t$ or $\sim m_b$ through the sum or difference
of two terms of roughly that order of magnitude.  We first consider the case
where $m_t$ comes from a single mass matrix element and $m_b$ also comes from a
(different) single matrix element (we will later need to relax the constraint
on $m_b$).

A simple mass matrix texture appropriate to our philosophy is:
\begin{eqnarray}
{\bf m_u } & = & \pmatrix{m_u&0&0\cr0&m_c&0\cr0&0&m_t\cr} \label{diagonalUp}\\
{\bf m_d } & = & \pmatrix{m_d&m_s V_{us}&m_b V_{ub}\cr0&m_s&m_b V_{cb}\cr0&0&m_b\cr} \approx {\bf V_{CKM}} \cdot \pmatrix{m_d&0&0\cr0&m_s&0\cr0&0&m_b\cr}\label{diagonalDown},
\end{eqnarray}
in which each small entry arises directly from the exponential in equation (\ref{massFromDisplacement}).  

\begin{figure}
\begin{center}
\leavevmode{\epsfysize=3.6truein \epsfbox{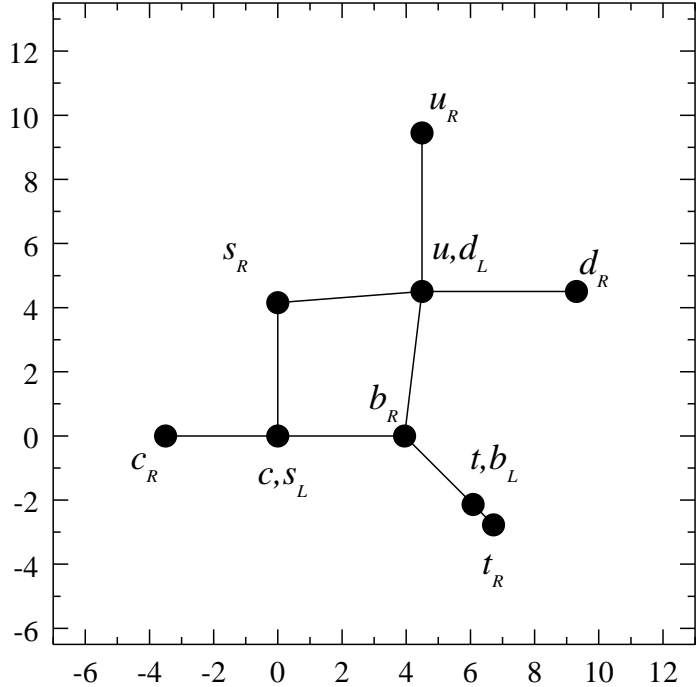}}
\end{center}
\caption{A possible set of locations for quark wavefunctions in a model with two extra dimensions. The lines represent constrained distances between wavefunction locations.}  
\label{twoDimensionalFigure}
\end{figure}

We now try to construct these matrices from a model.  Translating the matrices in Eqs. (\ref{diagonalUp}) and (\ref{diagonalDown}) into matrices of constrained relative distances using equation (\ref{displacementFromMass}) gives: 
\begin{eqnarray}
|q_i - u_j| & = & \mu^{-1} \pmatrix{4.8701 &(far)&(far)\cr (far)&3.4840 &(far)\cr (far)&(far)&.90052 \cr} \\
|q_i - d_j| & = & \mu^{-1} \pmatrix{4.7416 &4.4278  &4.5319 \cr (far)&4.0715 &3.9378 \cr (far)&(far)&3.0029 \cr}.
\end{eqnarray}
These constraints can easily be satisfied in two-dimensional extensions of the current model involving string defects in place of domain walls  (see figure \ref{twoDimensionalFigure}).  Unfortunately, the configuration cannot be collapsed down to one dimension without bringing some wavefunctions too close to preserve the smallness of the ``zero'' elements of the texture.

A similar texture can be found by taking the down mass matrix diagonal and using:
\begin{equation}
{\bf m_u } = \pmatrix{m_u&m_c V_{cd}&m_t V_{td}\cr0&m_c&m_t V_{ts} \cr 0 & 0 & m_t} \approx {\bf V^{\dagger}_{CKM}} \cdot \pmatrix{m_u&0&0\cr0&m_c&0\cr0&0&m_t\cr},
\end{equation}
and, at the same level of approximation, one can use textures where two of the $V_{CKM}$ elements come from one mass matrix and one comes from the other.  
The textures can be further extended by allowing the small masses to result from products of two small mass matrix entries by generalizing the observation that the $2 \times 2$ mass matrix:
\begin{equation}
\pmatrix{0&b\cr c&D}, 
\end{equation} 
with $b, c \ll D$ has masses $D$ and $bc/D$.  
For instance, the mass matrix:
\begin{equation}
{\bf m_d } =  \pmatrix{0&m_s V_{us}&m_b V_{ub}\cr m_d / V_{us}&m_s&m_b V_{cb}\cr0&0&m_b\cr}, 
\label{nonDiagonalTexture}
\end{equation}
leads to a down quark mass $m_d$. 

\section{Searching Parameter Space}
\label{searchingParameterSpace}

In this section we describe our efforts to find a set of locations for the
quark wavefunctions. The textures described in section
\ref{quarkSectorOfTheModel} (where each mixing parameter comes from a nonzero
above-diagonal element in either the up mass matrix or the down mass matrix,
and the masses come from either a diagonal element or a product of a
below-diagonal element and an above-diagonal mixing term) were searched
analytically with no success. The closest match allowed one to fit all the
masses and two of the mixings by hand, with all the ``zero'' elements of the
texture being of negligible magnitude.  Fitting these parameters determined all
the wavefunction positions, and thus predicted the remaining mixing
(parameterized by $V_{ub}$) which was found to be  many orders of magnitude too
small.   We then resorted to numerical methods, and also relaxed our
assumptions to allow for $m_t$  to result from the contributions of two or more
elements of roughly equal magnitude (and later extended this relaxation to
$m_b$).

We implemented a brute-force scan over the parameter space, which yielded a solution with two mass matrix elements of magnitude $m_b$ described in section \ref{analysisOfSolution} below.

\section{Analysis of Solution}
\label{analysisOfSolution}

\begin{figure}
\begin{center}
\leavevmode{\epsfxsize=6.00truein \epsfbox{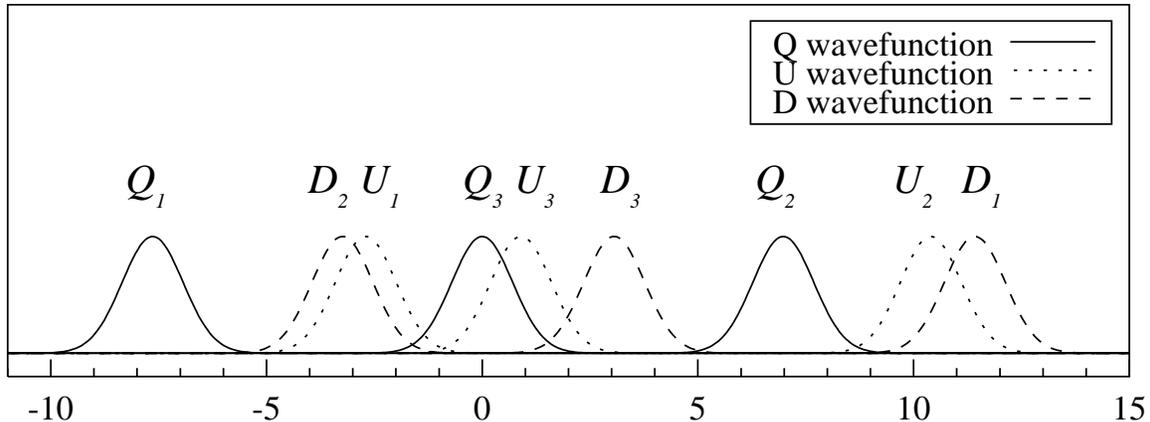}}
\end{center}
\caption{The locations of the quark wavefunctions.}
\label{oneDimensionalQuarkFigure}
\end{figure}

We now describe the configuration of quark wavefunction locations which was found to produce agreement with the observed masses and mixings of the standard model.  The configuration:
\begin{equation}
q_i  = \mu^{-1} \pmatrix{-7.6057 \cr 6.9522\cr 0.0}, \ \ \ \ \ \ \ u_i   =    \mu^{-1} \pmatrix{-2.7357\cr 10.4362 \cr 0.9012}, \ \ \ \ \ \ \ d_i    =   \mu^{-1} \pmatrix{11.3682\cr  -3.2250\cr3.0511 } 
\end{equation}
was found to produce mass matrices:\footnote{The element $m_{u(1,3)}= 5902.8 MeV$ is negligible because it is in the same row as the much larger top mass, and can be removed by a small 1-3 rotation redefining the $U_i^c$ fields.}
\begin{eqnarray}
{\bf m_u} & = & \pmatrix{1.7630  &  5.1637  \times 10^{-66} &  4.8087 \times 10^{-11}  \cr 1.0365 \times 10^{-15} & 576.06  &  2.7882 \times 10^{-3}  \cr 5902.8 & 5.5689  \times 10^{-19}& 165900  } MeV \\
& \approx & \pmatrix{m_u&0&0\cr0&m_c&0\cr0&0&m_t\cr} ,
\end{eqnarray} 
and:
\begin{eqnarray}
{\bf m_d} & = &\pmatrix{1.6660 \times 10^{-73} & 16.947 & 5.4422 \times 10^{-20} \cr 14.510  & 8.0344 \times 10^{-18} & 123.42   \cr 2.1526\times 10^{-23} & 1373.2 & 2370.2 } MeV \\
& \approx & \pmatrix{0&m_s V_{us}/{\rm cos}(\theta)&0\cr m_d / V_{us}&0&m_b V_{cb}/{\rm cos}(\theta) \cr0&m_b {\rm sin}(\theta)&m_b {\rm cos}(\theta)\cr} \label{quarkSolutionDownMassMatrix} .
\end{eqnarray}
which, using equations (\ref{firstComputePhysicalParameters}) - (\ref{lastComputePhysicalParameters}) and running using the scaling parameters in the appendix, gives masses:
\begin{equation}
\matrix{m_u =    3.24  \ MeV \cr m_c =  1.25  \ GeV \cr m_t =  166  \ GeV } \ \ \ \ \matrix{m_d =  6.01  \ MeV \cr m_s = 120 \ MeV \cr m_b =  4.25 \ GeV }
\end{equation}
and CKM matrix elements:
\begin{equation}
{\bf V_{CKM}} = \pmatrix{0.9755 & 0.2200 & 0.0031 \cr 0.2197  & 0.9748 & 0.0390 \cr 0.0116  & 0.0373 & 0.9992},
\end{equation}
which are consistent with experiment.  The configuration is depicted graphically in figure \ref{oneDimensionalQuarkFigure}.

The down mass matrix texture of equation (\ref{quarkSolutionDownMassMatrix}) differs from the example texture of equation (\ref{nonDiagonalTexture}) only by a 2-3 rotation by $\theta$ re-defining the $D_i^c$ fields, given that a certain additional relation among the entries holds.  
The rotation on the $D_i^c$ fields transforms the down mass matrix from 
equation (\ref{quarkSolutionDownMassMatrix}) as:
\begin{equation}
{\bf m_d } \rightarrow  {\bf m_d' }  = \pmatrix{0&m_s V_{us}&m_s V_{us} \ {\rm tan}(\theta)\cr m_d / V_{us}&m_b V_{cb} \ {\rm tan}(\theta)&m_b V_{cb}  \cr0& 0&m_b  \cr},
\end{equation}
which agrees with the texture of equation (\ref{nonDiagonalTexture}) if we can make: 
\begin{eqnarray}
m_s V_{us} \ {\rm tan}(\theta) & = & m_b V_{ub} \\
m_b V_{cb} \ {\rm tan}(\theta) & = & m_s \label{fixTheta}.
\end{eqnarray} 
Since $\theta$ is a free parameter we can always choose it to satisfy one of
the two equations, but the other will only be satisfied if
%
%
\begin{equation}
{m_b V_{ub} \over m_s V_{us}} = {\rm tan}(\theta) = {m_s \over  m_b V_{cb}} \label{relationOne}.
\end{equation} 
As the mass of the strange quark is the most uncertain of the quantities in equation (\ref{relationOne}), we re-write this relation as:
\begin{equation}
m_s = (V_{ub} V_{cb}/V_{us})^{1/2} \times m_b ,\label{relation}
\end{equation}
which holds at the common scale $m_t$.  Running the masses down to their physical scales using the factors in the appendix gives:
\begin{equation}
m_s^{\overline{MS}}(2 \  GeV) \approx (1.19) \times (V_{ub} V_{cb}/V_{us})^{1/2} \times m_b^{\overline{MS}}(m_b) \approx 120 \  MeV.
\label{lowMassRelation}
\end{equation}
The relation of equation (\ref{lowMassRelation}) is consistent with experiment, and represents the ``prediction'' present in this model which allows it to fit nine observables with eight model parameters.  

There are also seven other configurations giving the same masses and mixings
which result from the fact that, for all three $i$, moving the $U_i^c$ fields
to the opposite side of its partner quark doublet $Q_i$ while keeping their
relative distance fixed leaves the diagonal elements of ${\bf m}_u$ unchanged
while not carrying the $U_i^c$ so close to the other quark doublets $Q_{j \not=
i}$ that they create large off-diagonal elements in ${\bf m}_u$.  As these other configurations produce the same mass
matrices, they yield the same prediction for the strange mass of equation
(\ref{lowMassRelation}). 

\begin{figure}
\begin{center}
\leavevmode{\epsfxsize=6.00truein \epsfbox{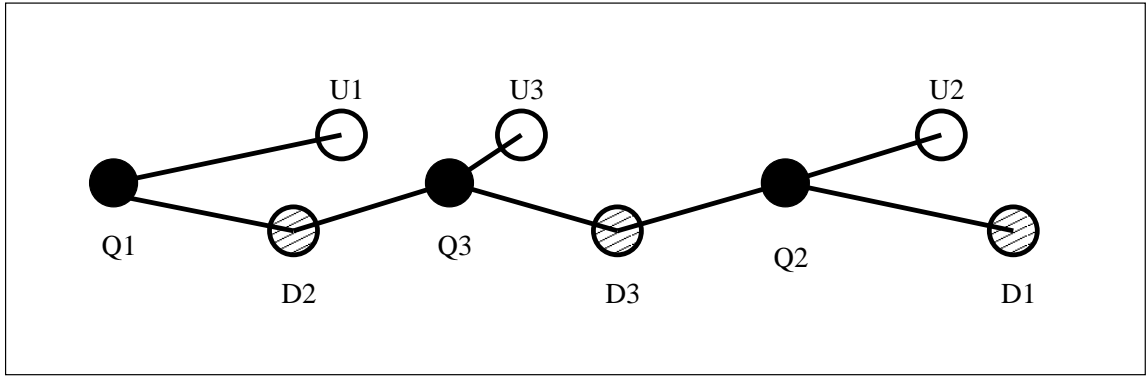}}
\end{center}
\caption{The constrained distances between the quark wavefunctions.}
\label{quarkMoleculeFigure}
\end{figure}

No other such distance preserving re-arrangements to the configuration can be
made without bringing otherwise distant fields too close and thus creating
large unwanted  mass matrix elements which destroy the predictions above.  The constrained distances between the wavefunction locations are illustrated in figure \ref{quarkMoleculeFigure}.

\section{Conclusion}
\label{gaussConclusion}

It is pleasing to find that Arkani-Hamed and Schmaltz's solution to the
Yukawa hierarchy problem can be made concrete, and that the masses and mixings
of the standard model, while highly constraining, do not have any hidden
relations that prevent their generation from a configuration of wavefunction
displacements.  The minimal model described contains the prediction
Eq. (\ref{lowMassRelation}) which may be falsified or supported by future
experimental data.

\bigskip

It is a pleasure to thank Nima Arkani-Hamed, Travis Brooks, Lance Dixon
and Michael Peskin for useful discussions and suggestions. This work
was supported by the Department of Energy under contract DE-AC03-76SF00515. 


\appendix

\section{Derivation of the Zero Mode}
\label{derivationOfZeroMode}

In this appendix we demonstrate that a minimal five-dimensional spinor coupled
to a scalar field with a generic domain-wall expectation value profile in the
fifth dimension leads to a single chiral fermion zero mode localized at the
zero crossing of the scalar field.  We show that if the scalar profile is
linear, the zero mode has a Gaussian profile, with the higher modes having the
spectrum and profile associated with harmonic oscillator wave-functions.  

The five-dimensional action:
\begin{equation}
S = \int d^5x \   \overline{\Psi} [i \gamma^A \partial_A  + \Phi(x_5)] \Psi,
\end{equation}
can be decomposed into four-dimensional terms:
\begin{equation}
S = \int d^4x \int dx_5   \ \overline{\Psi} [i \gamma^\mu \partial_\mu  + i \gamma^5 \partial_5  + \Phi  (x_5)] \Psi.
\end{equation}
In chiral spinor notation, this is:
\begin{equation}
S  = \int d^4x \int dx_5  \  \left[\matrix{\psi_L^\dagger & \psi_R^\dagger}
\right] \cdot \left[\matrix{i \partial_\mu \overline{\sigma}^\mu & -\partial_5 + \Phi(x_5) \cr +\partial_5 + \Phi(x_5) & i \partial_\mu \sigma^\mu}\right] \cdot \left[\matrix{\psi_L  \cr \psi_R}\right].
\end{equation}
We can split the spinors into a 4-dimensional spinor part $\psi$ multiplied by an extra-dimensional profile (with no spin indices) $\zeta$:
\begin{equation}
\Psi(x_m)  = \sum_i \left[\matrix{\psi_{L(i)}(x_\mu)\zeta_i^L(x_5) \cr \psi_{R(i)}(x_\mu)  \zeta_i^R(x_5)}\right],
\end{equation}
and integrate over $x_5$ to get:
\begin{eqnarray}
S & = &  \int d^4x \psi_{L(0)}^\dagger[i \overline{\sigma}_\mu \partial_\mu]\psi_{L(0)}  + \psi_{R(0)}^\dagger[i \sigma_\mu \partial_\mu]\psi_{R(0)} \nonumber \\
& & + \sum_i \overline{\Psi}_{(i)}[i \gamma_\mu \partial_\mu - m_i]\Psi_{(i)},
\end{eqnarray}
where the $m_i$ and $\zeta^{L,R}_i$ are chosen such that:
\begin{eqnarray}
(-\partial_5 + \Phi(x_5))\zeta^R_i(x_5) &=& m_i\zeta^L_i(x_5) \label{eigenA} \\
(+\partial_5 + \Phi(x_5))\zeta^L_i(x_5) &=& m_i\zeta^R_i(x_5) \label{eigenB},
\end{eqnarray}
and we have defined:
\begin{equation}
\Psi(x_\mu)_{(i)}  = \left[\matrix{\psi_{L(i)}(x_\mu)  \cr \psi_{R(i)}(x_\mu)}\right].
\end{equation}
From equations (\ref{eigenA}) and (\ref{eigenB}) we find zero modes:
\begin{eqnarray}
(-\partial_5 + \Phi(x_5))\zeta^R_0(x_5) & = & 0 \\
(+\partial_5 + \Phi(x_5))\zeta^L_0(x_5) & = & 0 ,
\end{eqnarray}
with solutions:
\begin{eqnarray}
\zeta^R_0(x_5) & = & A^R_0 e^{+\int_0^{x_5} \Phi(y_5) dy_5} \label{solutionA} \\
\zeta^L_0(x_5) & = & A^L_0 e^{-\int_0^{x_5} \Phi(y_5) dy_5}  \label{solutionB}, 
\end{eqnarray}
where the $A$ are normalization constants chosen to make $\int dx^5 \zeta^* \zeta = 1$.
In the approximation $\Phi(x_5) = 2 \mu^2 x_5$ this is:
\begin{eqnarray}
\zeta^R_0(x_5) & = & A e^{+\mu^2 x_5^2} \\
\zeta^L_0(x_5) & = & A e^{-\mu^2 x_5^2} ,
\end{eqnarray}
so the left-handed solution $\zeta^L_0(x_5)$ is a Gaussian and the right-handed solution $\zeta^R_0(x_5)$ is non-normalizable---only the left-handed solution is physical.  By inspection of equations (\ref{solutionA}) and (\ref{solutionB}) we see that this is a generic feature of any function $\Phi(x_5)$ that goes from a finite negative asymptotic value to a finite positive asymptotic value with one zero crossing.  If the Yukawa coupling were multiplied by $-1$ or if the scalar profile went from positive to negative, the right-handed solution would have been the physical solution and the left-handed solution would have been non-normalizable.

If the scalar expectation value dipped back through zero, as in a configuration of two domain walls coming from a kink/antikink profile, there would be a mirror zero-energy chiral fermion  stuck on the other wall.  In the limit that the second domain wall (and its mirror fermion) are infinitely far away, we recover the case discussed above.

To examine the massive spectrum, we operate on equation (\ref{eigenA}) with $(+\partial_5 + \Phi(x_5))$ and equation (\ref{eigenB}) with $(-\partial_5 + \Phi(x_5))$ and use $\Phi(x_5) = 2 \mu^2 x_5$ to obtain:
\begin{eqnarray}
(-\partial_5^2 + (2\mu^2 x_5)^2 + 2 \mu^2 )\zeta^R_i(x_5) &=& m^2_i\zeta^R_i(x_5) \label{squaredEigenA} \\
(-\partial_5^2 + (2\mu^2 x_5)^2 - 2 \mu^2)\zeta^L_i(x_5) &=& m^2_i\zeta^L_i(x_5) \label{squaredEigenB} .\\
\end{eqnarray}
Recognizing these as harmonic oscillator Schr\"odinger equations for a particle with mass $\hbar^2/2$, characteristic angular frequency $\omega_0 = 4 \mu^2/\hbar$ and ground state energy $\pm 2 \mu^2$, we can immediately write down the solution:
\begin{eqnarray} 
m_n^2 & = & 4 \mu^2 n \ \ \ (n = 0,1,2,3,...) \\
\zeta^L_n(x_5) =  \zeta^R_{n+1}(x_5)  & = &  ( {a^\dagger \over \sqrt{n!}})^n\zeta^L_0(x_5) \equiv {1 \over \sqrt{n!}}({-\partial_5 + 2 \mu^2 x_5 \over 2 \mu})^n\zeta^L_0(x_5) .
\end{eqnarray}  
In this framework we can see the chirality of the zero mode by noting that the ground state energy shift term $\pm 2 \mu^2$ of the Schr\"odinger equation shifts the eigenvalue tower from the usual $\sim1/2,3/2,5/2...$ down to $\sim0,1,2...$ for the left handed modes and up to $\sim1,2,3...$ for the right handed modes.  

\section{Standard Model Parameters}
\label{standardModelParameters}

For the experimentally allowed values of the standard model parameters, we used\cite{pdb}:
\begin{eqnarray}
m_u & = & 1.5 \ {\rm to}  \ 5  \ MeV \nonumber \\ 
m_d & = & 3  \ {\rm to} \  9  \ MeV \nonumber \\ 
m_s & = & 60  \ {\rm to}  \ 170  \ MeV \nonumber \\ 
\nonumber \\
m_u/m_d & = & 0.20  \ {\rm to}  \ 0.70 \nonumber \\ 
(m_u + m_d)/2 & = & 2  \ {\rm to}  \ 6  \ MeV  \nonumber \\ 
{m_s - (m_u + m_d)/2\over m_d - m_u} & = & 34 \ {\rm to}  \ 51 \hskip1.4in \ \nonumber \\
\nonumber \\
m_c & = & 1100  \ {\rm to}  \ 1400  \ MeV \nonumber \\ 
m_b & = & 4100  \ {\rm to} \  4400 \  MeV \nonumber \\ 
m_t & = & 166000 \pm 5000  \ MeV \nonumber \\  
\nonumber \\
V_{us} & = & 0.217  \ {\rm to} \  0.224 \nonumber \\ 
V_{ub} & = & 0.0018 \  {\rm to}  \ 0.0045 \nonumber \\ 
V_{cb} & = & 0.036 \  {\rm to}  \ 0.042 \nonumber \\
\nonumber \\
m_e & = & 0.5110 \ MeV \nonumber \\ 
m_{\mu} & = & 105.7   MeV \nonumber \\ 
m_{\tau} & = & 1777  \ MeV  
\end{eqnarray}
where the quark masses are in the $\overline{MS}$ renormalization scheme.  The up, down, and strange masses are evaluated at a scale of $2 \ GeV$, while the other quark masses are evaluated at a scale equal to their $\overline{MS}$ mass.  The lepton masses are pole masses. 

To run the quark masses to the common scale $m_t$, we used the three-loop QCD and one-loop QED scaling factors\cite{babuMohapatra}:
\begin{equation}
\begin{tabular}{|c|c|c|c|c|c|}                     
\hline
$\eta_u$ & $\eta_d$ &  $\eta_s$ & $\eta_c$ & $\eta_b$ & $\eta_t$ \\ \hline
1.84 & 1.84 & 1.84 & 2.17 & 1.55 & 1.00 \\ \hline
\end{tabular} 
\end{equation} 
where $\eta_i = m_i(m_i)/m_i(m_t)$ for $i=c,b,t$ and $\eta_i = m_i(2 \ GeV)/m_i(m_t)$ for $i=u,d,s$.  The lepton pole masses were related to their $\overline{MS}$ masses evaluated at $m_t$ using the relation 
\begin{equation}
m^{\rm pole} = m^{\overline{MS}}(\mu) \ \left(1 + {\alpha \over \pi} \left[1 +  {3 \over 4} {\rm log}({\mu^2 \over m^2}) \right] \right)
\end{equation}
to give scaling factors
\begin{equation}
\begin{tabular}{|c|c|c|}                     
\hline
$\eta_e$ & $\eta_\mu$ &  $\eta_\tau$ \\ \hline
1.05 & 1.03 & 1.02 \\ \hline
\end{tabular} 
\end{equation} 
where $\eta_i = m^{\rm pole}_i(m_i)/m^{\overline{MS}}_i(m_t)$.

%


\begin{thebibliography}{99}




\bibitem{arkaniHamedSchmaltz}
	N. Arkani-Hamed and M. Schmaltz, hep-ph/9903417. 

\bibitem{flavor}
	N. Arkani-Hamed, L.J. Hall, D. Smith and N. Weiner, hep-ph/9909326; 
	N. Arkani-Hamed and S. Dimopoulos, hep-ph/9811353.

\bibitem{KK}
        I. Antoniadis, K. Benakli and M.Quiros, \Journal{\PLB}{331}{313}{1994};
	P. Nath and M. Yamaguchi, hep-ph/9902323 and hep-ph/9903298;
	M. Masip and A. Pomarol, hep-ph/9902467; 
	W.J. Marciano, hep-ph/9903451; 
	L.J. Hall and C. Kolda, \Journal{\PLB}{459}{213}{1999}; 
	R. Casalbuoni, S. DeCurtis and D. Dominici, hep-ph/9905568;
	R. Casalbuoni, S. DeCurtis, D. Dominici and R. Gatto, hep-ph/9907355;
	A. Strumia, hep-ph/9906266; 
	C.D. Carone, hep-ph/9907362;
	T.G. Rizzo and 	J.D. Wells, hep-ph/9906234.
	
\bibitem{electroweakData}
	T.G. Rizzo, hep-ph/9909232; See also 
        I. Antoniadis, \Journal{\PLB}{246}{377}{1990};
	I. Antoniadis and K. Benakli, \Journal{\PLB}{326}{69}{1994};
        I. Antoniadis, K. Benakli and M. Quiros, hep-ph/9905311;
	P. Nath, Y. Yamada and M. Yamaguchi, hep-ph/9905415.

\bibitem{yuval}
	N. Arkani-Hamed, Y. Grossman, M. Schmaltz, hep-ph/9909411. 

\bibitem{FCNC}
	A. Delgado, A. Pomarol, M. Quiros hep-ph/9911252.
	
\bibitem{pdb}
	C. Caso et.al. {\it Particle Data Group}, 
	\Journal{\EPJC}{3}{1}{1998}.
	
\bibitem{babuMohapatra}
	V. Barger, M.S. Berger, T. Han and M. Zralek,
	\Journal{\PRL}{68}{3394}{1992},
	hep-ph/9203220.

	G.W. Anderson, S. Raby, S. Dimopoulos, L.J. Hall,
	\Journal{\PRD}{47}{3702}{1993},
	hep-ph/9209250.
	

\end{thebibliography}
\end{document}